\documentclass[runningheads]{llncs}

\usepackage{makeidx}  
\usepackage{graphicx}
\usepackage{subfigure}
\usepackage{multirow}
\usepackage{color,soul} 
\usepackage{ dsfont }
\usepackage{amssymb,amsfonts}
\usepackage{physics} 
\usepackage{cases}
\usepackage{hyperref}
\usepackage{multirow}
\usepackage{amsmath}

\usepackage{graphicx}

\usepackage{amsmath}

\usepackage{subfigure}
\usepackage{algorithm}
\usepackage{algorithmic}

\newcommand{\be}{\begin{equation}}
\newcommand{\ee}{\end{equation}}

\makeatletter
 \def\SOUL@hlpreamble{%
 \setul{}{2.4ex}
 \let\SOUL@stcolor\SOUL@hlcolor
 \SOUL@stpreamble
 }
\makeatother

\begin{document}
\mainmatter              

\title{CLTS-GAN: Color-Lighting-Texture-Specular Reflection Augmentation for Colonoscopy}
\titlerunning{CLTS-GAN}  
%

\author{Shawn Mathew$^1$* \and Saad Nadeem$^2$*{\let\thefootnote\relax\footnote{{\hspace{-4mm}*Equal contribution. [shawmathew,ari]@cs.stonybrook.edu and nadeems@mskcc.org}}} \and Arie Kaufman$^1$}
\index{Mathew, Shawn}
\index{Nadeem, Saad}
\index{Kaufman, Arie}

\authorrunning{Mathew \emph{et al}.} 

\tocauthor{Shawn Mathew, Saad Nadeem, and Arie Kaufman}

\institute{$^1$Department of Computer Science, Stony Brook University\\
$^2$Department of Medical Physics, Memorial Sloan Kettering Cancer Center\\
}

\maketitle              

\begin{abstract}
{
Automated analysis of optical colonoscopy (OC) video frames (to assist endoscopists during OC) is challenging due to variations in color, lighting, texture, and specular reflections. Previous methods either remove some of these variations via preprocessing (making pipelines cumbersome) or add diverse training data with annotations (but expensive and time-consuming). We present CLTS-GAN, a new deep learning model that gives fine control over color, lighting, texture, and specular reflection synthesis for OC video frames. We show that adding these colonoscopy-specific augmentations to the training data can improve state-of-the-art polyp detection/segmentation methods as well as drive next generation of OC simulators for training medical students. The code and pre-trained models for CLTS-GAN are available on Computational Endoscopy Platform GitHub (\url{https://github.com/nadeemlab/CEP}).

}

\end{abstract}
\section{Introduction}
Colorectal cancer is the fourth deadliest cancer. Polyps, anomalous protrusions on the colon wall, are precursors of colon cancer and are often screened and removed using optical colonoscopy (OC). During OC, variations in color, texture, lighting, specular reflections, and fluid motion make polyp detection by a gastroenterologist or an automated method challenging. Previous methods deal with these variations either by removing specular reflections \cite{ma2019real,ma2021rnnslam}, removing color/texture \cite{mahmood2018unsupervised}, and correcting lighting \cite{zhang2021lighting} in the preprocessing steps (making pipelines cumbersome) or by adding more diverse training data with expert annotations (but expensive and time-consuming). If the automated methods can be made invariant to color, lighting, texture, and specular reflections without adding any preprocessing overhead or additional annotations, then these methods can act as effective second readers to gastroenterologists, improving the overall polyp detection accuracy and potentially reducing the procedure time (end-to-end colon wall inspection from rectum to cecum and back).

We present a new deep learning model, CLTS-GAN, that provides fine-grained control over creation of colonoscopy-specific color, lighting, texture, and specular reflection augmentations. Specifically, we use a one-to-many image-to-image translation model with Adaptive Instance Normalization (AdaIn) and noise input (StyleGAN \cite{karras2019style}) to create these augmentations. Color and lighting augmentations are performed by injecting 1D vectors (sampled from a uniform distribution) using AdaIn, while texture and specular reflection augmentations are incorporated by directly adding 2D matrices (sampled from a uniform distribution) to the latent features. The color and lighting vectors can be extracted from one OC image and used to modify the color and lighting of another OC image. We show that these colonoscopy-specific augmentations to the training data can improve accuracy of the state-of-the-art deep learning polyp detection methods as well as drive next generation OC simulators for teaching medical students \cite{fazlollahi2022effect}. The contributions of this work are as follows:
\begin{enumerate}
    \item  CLTS-GAN, an unsupervised one-to-many image-to-image translation model
    \item A novel texture loss to encourage a larger variety in texture and specular generation for OC images
    \item A method for augmenting colonoscopy frames that produces state-of-the-art results for polyp detection
    \item Latent space analysis to make CLTS-GAN more interpretable for generating color, lighting, texture, and specular reflection 
\end{enumerate}

\section{Related Works}
The image-to-image translation task aims to translate an image from one domain to another. Certain applications have access to ground truth information providing supervision for models like pix2pix \cite{isola2017image}. Zhu et al. developed CycleGAN, an image-to-image translation model without needing ground truth correspondence  \cite{chu2017cyclegan}. This is done using a cycle consistency loss that drives other unsupervised domain translation models. Examples include MUNIT\cite{huang2018multimodal} and Augmented CycleGAN\cite{almahairi2018augmented} which additionally incorporated noise to learn a many-to-many domain translation. This many-to-many mapping lacks control over specific image attributes. XDCycleGAN \cite{mathew2020augmenting} and FoldIt \cite{mathew2021foldit} model one-to-many image-to-image translation, however their networks functionally learn a one-to-one mapping.

Generating realistic OC from CT scans has been used for OC simulators. VRCaps uses a rendering approach to simulate a camera inside organs captured in CT scans \cite{incetan2021vr}. For the colon, a simple texture is mapped on a mesh where OC artifacts (e.g., specular reflections, fish-eye lense distortion) are added. However, it cannot produce complex textures and colors normally found in OC. OfGAN uses image-to-image translation with optical flow to transform colon simulator images to OC \cite{xu2020ofgan}. It uses synthetic colonoscopy frames embedded with texture and specular reflection, which improve the realism of generated images. The texture and specular mapping in the synthetic frames, however, restrict additional texture and specular generation. Rivoir et al. use neural textures to create realistic and temporally consistent textures \cite{rivoir2021long}. They require a full 3D mesh to embed the neural textures making it difficult to augment annotated real videos.

\section{Data}

\begin{figure}[t!]
\centering
\includegraphics[width=.85\textwidth]{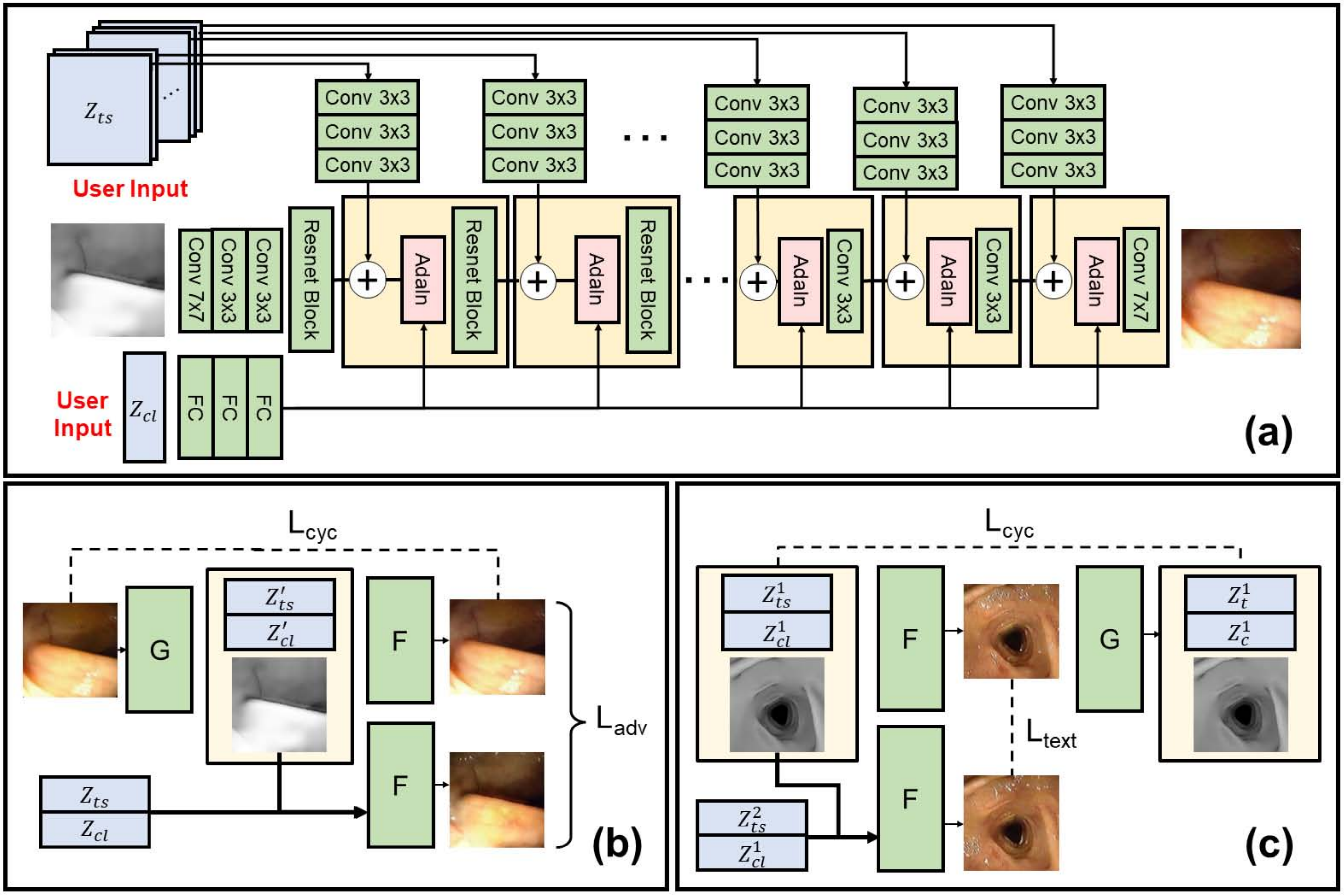}
\caption{ (a) shows user specified noise being used in $F$. $z_{ts}$ is a set of 2D matrices that goes through convolutional layers and is added to latent features throughout the network. $z_{cl}$ is a 1D vector that goes through fully connected layers and is distributed to AdaIn layers. Both $z_{ts}$ and $z_{cls}$ are sampled from a uniform distribution and can be sampled until the user is satisfied with the result. (b) depicts the forward cycle where OC passes through $G$, predicting its noise vectors and VC. These are then passed into $F$ to reconstruct the image. $F$ produces another OC image using different noise vectors where $\mathcal{L}_{adv}$ is applied. (c) depicts the backwards cycle where a VC image with different $Z_{ts}$ is passed into $F$. The resulting two OC images have $\mathcal{L}_{text}$ applied. One OC image is used for reconstruction via $G$ where $\mathcal{L}_{cyc}$ is applied.}
\label{fig:network}
\end{figure}

10 OC videos and 10 abdominal CT scans for virtual colonoscopy (VC) were obtained at Stony Brook University Hospital. The OC videos were rescaled to 256x256 and cropped to remove borders. Since the colon is deformable and CT scans capture a single time point, there is no ground truth correspondence between OC and VC. The VC data uses triangulated meshes from abdominal CT scans similar to \cite{nadeem2016computer}. Flythroughs were generated using Blender with two lights on both sides of the camera to replicate a colonoscope. Additionally, the inverse square fall-off property was applied to accurately simulate lighting conditions in OC. A total of 3000 VC and OC frames were extracted. 1500 were used for training while  900 and 600 were used for validation and testing.

\section{Methods}


CLTS-GAN  is composed of two generators and three discriminators. One generator, $G$, uses OC to predict VC with two corresponding noise parameters. The first parameter, $z_{ts}$, is a number of matrices that represent texture and specular reflection information. The second parameter, $z_{cl}$, is a 1D vector that contains color and lighting information. The second generator, $F$, uses $z_{ts}$ and $z_{cl}$ to transform a VC image into a realistic OC image.  Figure \ref{fig:network}a shows how the noise values are used in $F$. $z_{cl}$ is incorporated using AdaIn layers, which globally affects the latent features. $z_{ts}$ is directly added to latent features offering localized information. The complete objective function for the network is defined as:

\begin{equation}
     \mathcal{L}_{obj} =  \lambda_{adv}\mathcal{L}_{adv} +  \lambda_{cyc}\mathcal{L}_{cyc} +  \lambda_{t}\mathcal{L}_{t}
     +\lambda_{idt}\mathcal{L}_{idt}
\end{equation}

Cycle consistency is used in many image-to-image translation models and ensures features from the input are present in the output when transformed. The cycle consistency loss used for OC is shown in Figure \ref{fig:network}b and defined as:

\begin{equation}
    \mathcal{L}_{cyc}^{OC}(G,F,A) = \mathds{E}_{x \backsim p(A)} \|x - F(G_{im}(x),G_{cl}(x),G_{ts}(x))\|_1
\end{equation}
where $x \backsim p(A)$ represents a data distribution and $G_{im}$, $G_{cl}$ and $G_{ts}$ represents $G$'s output. Since $G$ has additional outputs, the cycle consistency loss should incorporate these extra vectors as seen in Figure \ref{fig:network}c.
 
\begin{equation}
\begin{split}
    \mathcal{L}_{cyc}^{VC}(G,F,A,Z) = \mathds{E}_{x \backsim p(A),z \backsim p(Z)} & \|x - G_{im}(F(x,z_{cl},z_{ts}))\|_1 + \\
    & \|z_{cl} - G_{cl}(F(x,z_{cl},z_{ts}))\|_1 +\\
    & \|z_{ts} - G_{ts}(F(x,z_{cl},z_{ts}))\|_1 
\end{split}
\end{equation}

The cycle consistency component of the objective loss function is defined as:

\begin{equation}
\begin{split}
    \mathcal{L}_{cyc} = \mathcal{L}_{cyc}^{OC}(G,F,OC) + \mathcal{L}_{cyc}^{VC}(G,F,VC,Z)
\end{split}
\end{equation}

Each generator has a discriminator, $D$, which adds an adversarial loss so the output  resembles the output domain. The adversarial loss for each GAN is:
\begin{equation}
     \mathcal{L}_{GAN}(G,D,A,B) =  \mathds{E}_{y \backsim p(B)} \big[ \textrm{log} (D(y))\big] + \\
     \mathds{E}_{x \backsim p(A)} \big[\textrm{log}(1 - D(G(x))\big],
\end{equation}
 $G$ has noise vectors in its output so an additional discriminator is required. Rather than distinguishing noise values, a discriminator is applied to recreated images since our concern lies with the imaging rather than the noise. The discriminator compares images produced by random noise vectors and vectors produced by F. This adversarial loss is shown in Figure \ref{fig:network}b and is defined as:
\begin{equation}
\begin{split}
     \mathcal{L}_{GAN}^{rec}(G,F,D,A) = & \mathds{E}_{x \backsim p(A)} \big[ \textrm{log} (D(F(G_{im}(x),G_{cl}(x),G_{ts}(x)))\big] + \\
     & \mathds{E}_{x \backsim p(A), z \backsim p(Z)} \big[\textrm{log}(1 - D(F(G_{im}(x),z_{cl},z_{ts})))\big],
\end{split}
\end{equation}

The adversarial portion of the objective loss is as follows:
\begin{equation}
\begin{split}
     \mathcal{L}_{adv} = \ & \mathcal{L}_{GAN}(G,D_{G},OC,VC) + \mathcal{L}_{GAN}(F,D_{F},VC,OC) + \\        
     & \mathcal{L}_{GAN}^{rec}(G,F,D_{rec},OC) 
\end{split}
\end{equation}

\begin{figure}[t!]
\begin{center}

\includegraphics[width=.9\textwidth]{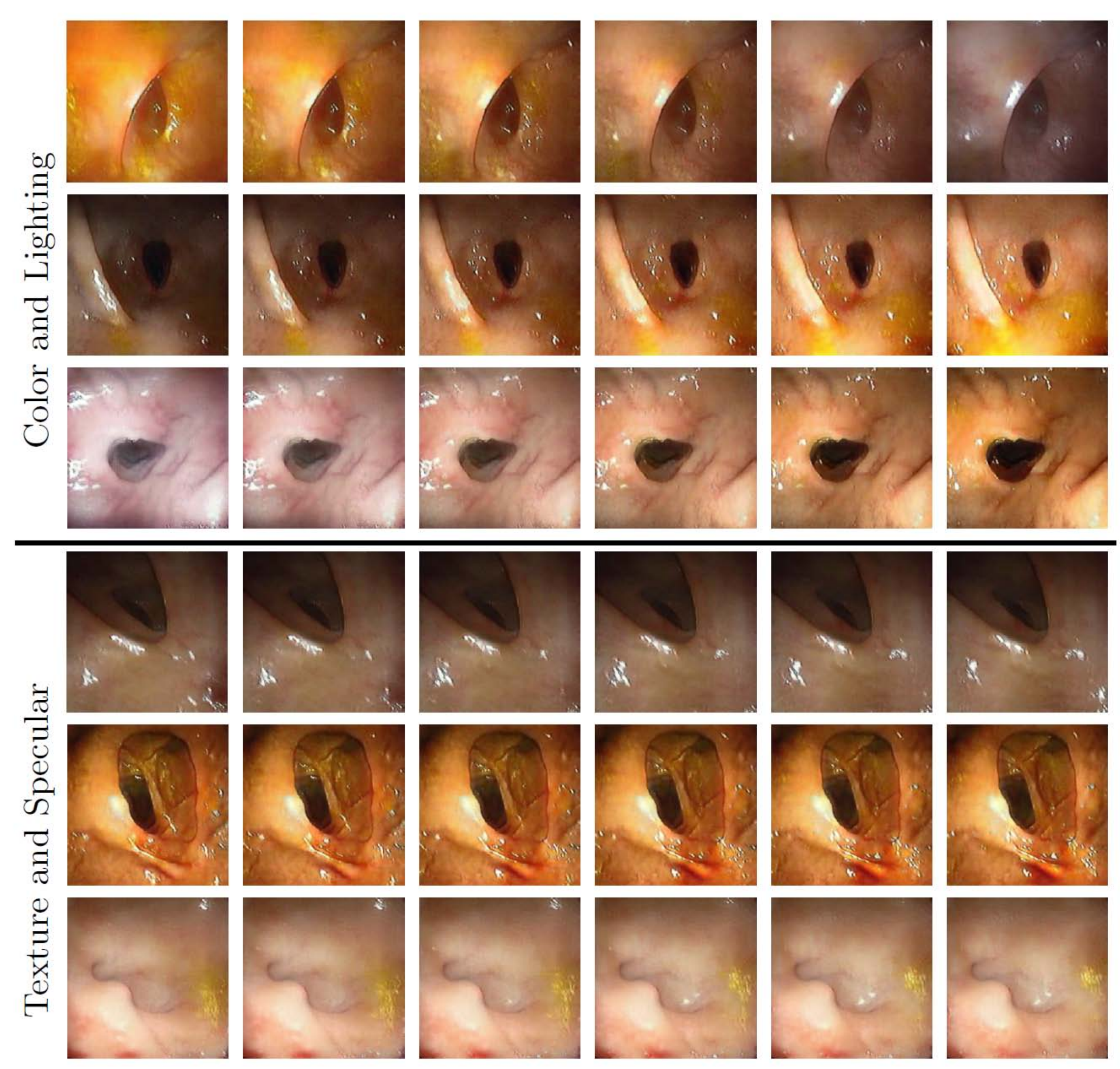}

\caption{ To understand how $z_{cl}$ and $z_{ts}$ affect the output, $z_{cl}$ and $z_{ts}$ are individually linearly interpolated. The top half shows interpolation between $z_{cl}$ values, while $z_{ts}$ is fixed.  The colon-specific color and lighting gradually changes with $z_{cl}$.  The bottom half shows $z_{cl}$ fixed, while $z_{ts}$ is interpolated. The specular reflection shapes and texture gradually change. The last row also shows fecal matter changing between images.}
\vspace{-1em}
\label{fig:walk}
\end{center}
\end{figure}

During training, $F$ may ignore $z_{ts}$. To encourage using noise input, $\mathcal{L}_{t}$ is added to penalize the network when different noise inputs have similar results. The function penalizing the network is defined as:

\begin{equation*}
\mathcal{L}_{text}(I_1,I_2) =  \begin{cases}
          \alpha - \|I_1 - I_2\|_1 \quad &\text{if } \alpha > \|I_1 - I_2\|_1 \,  \\
          0 \quad &\text{else}  \\
     \end{cases}
\end{equation*}
where $I$ is an image and $\alpha$ they differ.
$F$ is applied to two different images, and the OC images are compared using $L_{t}$ as seen in Figure \ref{fig:network}c and defined as:

\begin{equation*}
\begin{split}
\mathcal{L}_{t} =   \mathds{E}_{x \backsim p(VC), z \backsim p(Z)} & \mathcal{L}_{text}(F(x,z_{cl},z_{ts}^1), F(x,z_{cl},z_{ts}^2)) \\
\end{split}
\end{equation*}

Lastly, an identity loss is added for stability. An image should be unchanged if the input is from the output domain. It is only applied to $G$ to encourage texture and specular reflection generation. The identity loss is defined as:
\begin{equation}
  \mathcal{L}_{idt}(G,A) = \mathds{E}_{x \backsim p(A)} \|x - G_{im}(x) \|_1
\end{equation}

The identity portion of the objective loss is defined as $\mathcal{L}_{idt} = \mathcal{L}_{idt}(G,VC)$. The generators are ResNets \cite{he2016deep} with 9 blocks that use 23MB. CLTS-GAN uses PatchGAN discriminators \cite{isola2017image}, each using 3MB. The network was trained for 200 epochs on an Nvidia RTX 6000 GPU with the following parameters: $\lambda_{adv} = 1, \lambda_{T} = 10, \lambda_{text} = 20, \lambda_{idt} = 1$, and $\alpha = .1$. Inference time is .04 seconds.

\begin{figure}[t!]
\begin{center}

\includegraphics[width=.8\textwidth]{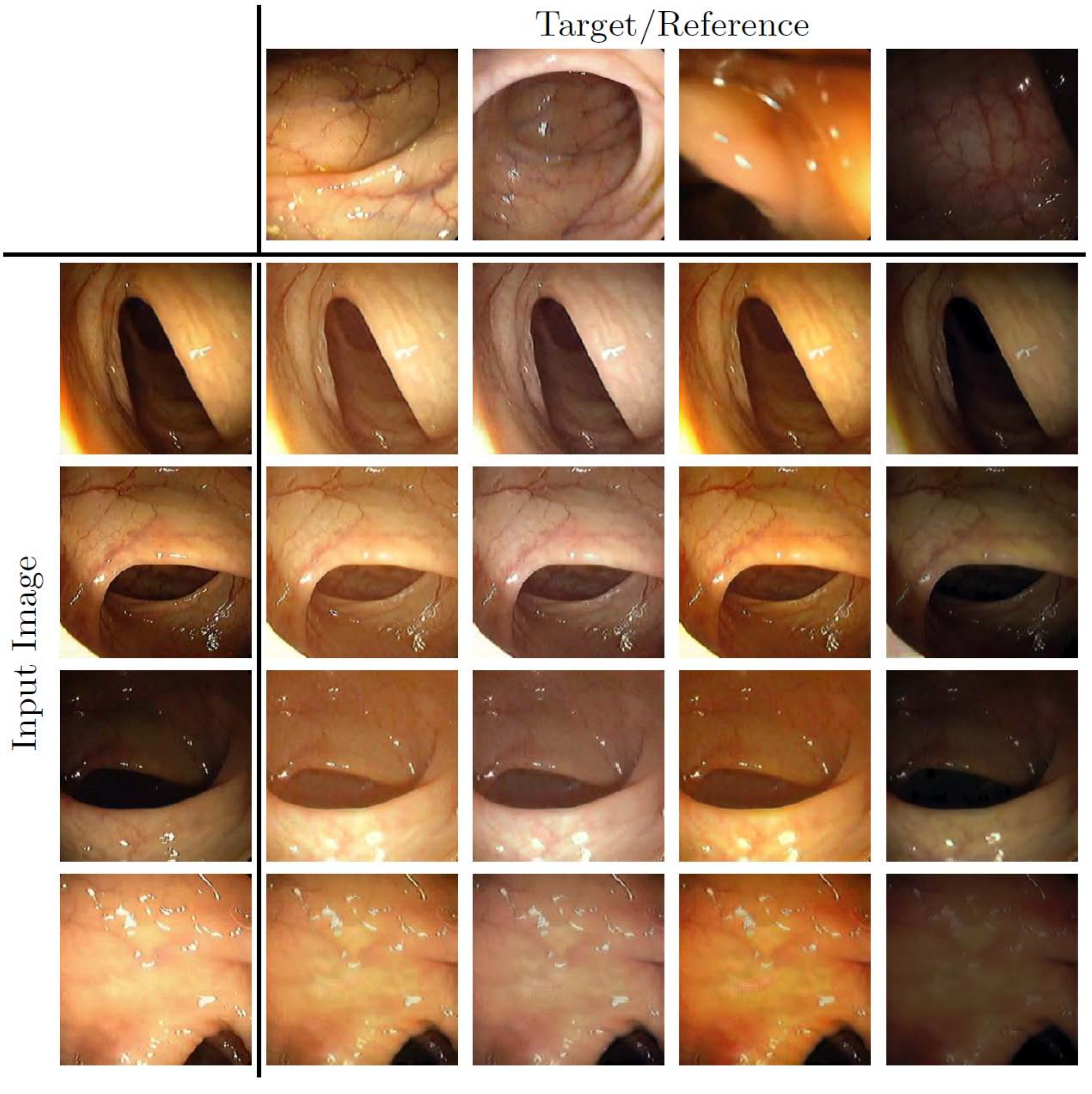}

\caption{Showing the $z_{cl}$ vector being extracted from various reference images (top most) and applied to target images (left most) to transfer its colon-specific color and lighting. }
\vspace{-1em}
\label{fig:transfer}
\end{center}
\end{figure}

CLTS-GAN controls the output using $z_{ts}$ and $z_{cl}$. For VC, if two $z_{cl}$ values are selected with a fixed $z_{ts}$ they can be linearly interpolated and passed into $F$ creating gradual changes in the colon-specific color and lighting as seen in Figure \ref{fig:walk}.  The strength of the specular reflections change with $z_{cl}$ since the lighting is being altered. Similarly, $z_{ts}$ can be linearly interpolated to provide gradual changes in texture and specular reflection as well as fecal matter. Here the shape of the specular reflections and texture fade in and out. Since changes in $z_{ts}$ and $z_{cl}$ do not lead to sporadic changes, they can be used in more meaningful ways.

\begin{figure}[t!]
\begin{center}

\includegraphics[width=.9\textwidth]{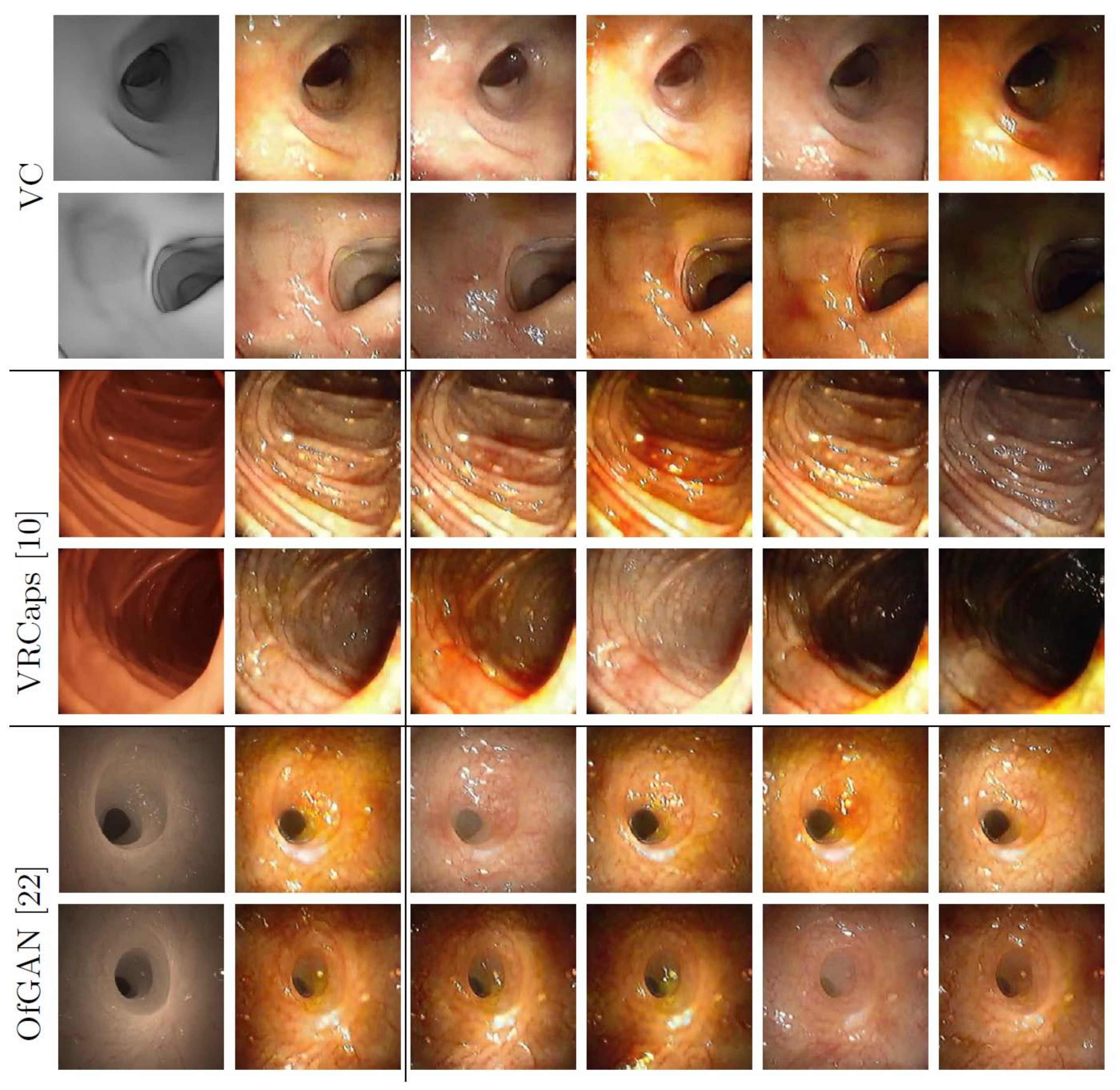}

\caption{Depicting our model using various $z_{cl}$ and $z_{ts}$  values to generate realistic OC images. The left most image is the input image for CLTS-GAN followed by the output OC images. We show results on VC, VRCaps \cite{incetan2021vr} data, and OfGAN \cite{xu2020ofgan} synthetic input. Additional results can be found in Figure 1 of the supplementary material.}
\vspace{-1em}
\label{fig:allrand}
\end{center}
\end{figure}

Figure \ref{fig:transfer} shows the transfer of colon-specific color and lighting information from one OC image to another.  $G$ extracts the $z_{cl}$ vector from the reference and the VC and $z_{ts}$ from the target. When these values are input to $F$ it transfers the color and lighting from the reference to the target. $z_{ts}$ remains fixed since it is intended for generating realistic textures and specular for VC instead of altering geometry dependant texture and specular of OC.

\section{Results and Discussion}

Figure \ref{fig:allrand} shows qualitative results for CLTS-GAN's realistic OC generation using VC images and data from VRCaps \cite{incetan2021vr} and OfGAN \cite{xu2020ofgan}. The input was passed to $F$ with $z_{ts}$ and $z_{cl}$ randomly sampled from a uniform distribution to show a large variety in colon-specific color, lighting, texture and specular reflection. More results can be found in the supplementary material. $z_{ts}$ and $z_{cl}$ can be individually changed to control the texture and specular reflection separately from the color and lighting as shown in Figures 2 and 3 of the supplementary material.

\begin{figure}[t!]
\begin{center}

\includegraphics[width=.8\textwidth]{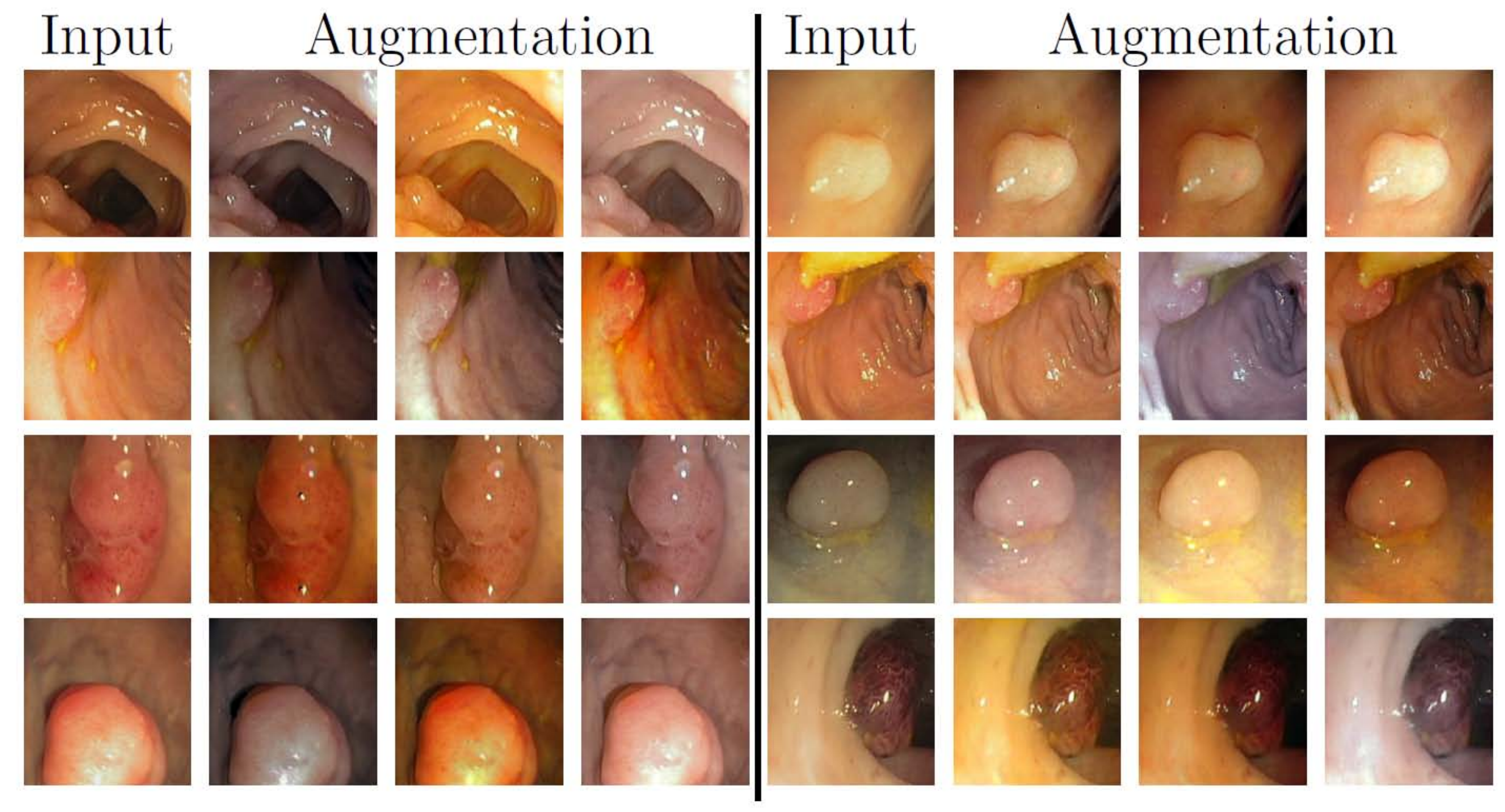}

\caption{Augmented data from CVC Clinic DB \cite{bernal2015wm}. The images go through $G$ to extract VC and $z_{ts}$. $z_{cl}$ is sampled from a uniform distribution and passed into $F$.}
\vspace{-1em}
\label{fig:polyp}
\end{center}
\end{figure}

To show quantitative evaluation of CLTS-GAN, PraNet \cite{fan2020pranet}, a state-of-the-art polyp segmentation model, is trained with and without augmentation. PraNet uses CVC Clinic DB \cite{bernal2015wm} and HyperKvasir \cite{borgli2020} for training. The images were augmented with colon-specific color and lighting, while polyp specific textures and speculars were preserved. Random $z_{cl}$ values are applied to training images by extracting the VC and $z_{ts}$ using $G$ and passing the three values to $F$. Examples are shown in Figure \ref{fig:polyp}. PraNet was trained having each image augmented 0, 1, and 3 times. When there was no augmentation or one augmentation the network was trained for 20 epochs. To avoid overfitting on the shapes of the polyps, the network was trained for 10 epochs when augmented 3 times. Testing results are shown in Table \ref{tab:Pranet}. Data augmentation from CLTS-GAN improves the DICE, IoU, and MAE scores for various testing datasets. For the CVC-T dataset, using only one augmentation appeared to have marginal improvement over using 3.

\begin{table}[]
\setlength{\tabcolsep}{2pt}
\centering
\caption{PraNet results with and without dataset augmentation. Colon-specific color and lighting augmentation was applied to avoid altering polyp specific textures. Results for 1 and 3 additional images are shown in the second and third rows. Both show improvement over PraNet without augmentation. PraNet with 1 augmentation is better for CVC-T which indicates the network may have overfit on the shapes of polyps.}
\label{tab:Pranet}
\begin{tabular}{l|ccc|ccc|ccc}

                 & \multicolumn{3}{c|}{CVC-Colon DB \cite{bernal2012towards}} & \multicolumn{3}{c|}{ETIS \cite{silva2014toward}} & \multicolumn{3}{c}{CVC-T \cite{vazquez2017benchmark}} \\ \hline
                 & Dice$\uparrow$      & IoU$\uparrow$       & MAE$\downarrow$       & Dice$\uparrow$    & IoU$\uparrow$    & MAE$\downarrow$    & Dice$\uparrow$    & IoU$\uparrow$    & MAE$\downarrow$    \\
PraNet w/out Aug & 0.712     & 0.640     & 0.043     & 0.628   & 0.567  & 0.031  & 0.871   & 0.797  & 0.10   \\
PraNet w/ 1 Aug    & 0.750 & 0.671 & 0.037 & 0.704 & 0.626 & \textbf{0.019} & \textbf{0.893} & \textbf{0.824} & \textbf{0.007} \\
PraNet w/ 3 Aug    & \textbf{0.781} & \textbf{0.697} & \textbf{0.030} & \textbf{0.710} & \textbf{0.639} & 0.027 & 0.884 & 0.815 & 0.010

\end{tabular}
\end{table}

In this work we present CLTS-GAN, a one-to-many image-to-image translation model for dataset augmentation and OC synthesis with control over color, lighting, texture, and specular reflections. $z_{ts}$ and $z_{cl}$ control these attributes, but can be further disentangled. High intensity specular reflections can be extracted with a loss and stored in a separate parameter. CLTS-GAN does not contain temporal components. Adding multiple frames as input can get the network to use the texture and specular information in a temporally consistent manner. Moreover, in the future, we will also explore the utility of CLTS-GAN augmentations in depth inference \cite{mahmood2018unsupervised,mathew2020augmenting} and folds detection \cite{mathew2021foldit}. We hypothesize that the full gamut of color-lighting-texture-specular augmentations can be used in these scenarios to improve performance.

\section*{Acknowledgements}
This project was supported by MSK Cancer Center Support Grant/Core Grant (P30 CA008748) and NSF grants CNS1650499, OAC1919752, and ICER1940302.

\title{CLTS-GAN: Color-Lighting-Texture-Specular Reflection Augmentation for Colonoscopy (Supplement)}
\titlerunning{CLTS-GAN}  
%

\author{Shawn Mathew$^1$* \and Saad Nadeem$^2$*{\let\thefootnote\relax\footnote{{\hspace{-4mm}*Equal contribution}}} \and Arie Kaufman$^1$}
\index{Mathew, Shawn}
\index{Nadeem, Saad}
\index{Kaufman, Arie}

\authorrunning{Mathew \emph{et al}.} 

\tocauthor{Shawn Mathew, Saad Nadeem, and Arie Kaufman}

\institute{$^1$Department of Computer Science, Stony Brook University\\
$^2$Department of Medical Physics, Memorial Sloan Kettering Cancer Center\\
}

\maketitle              

\begin{figure}[!]
\begin{center}

\includegraphics[width=.9\textwidth]{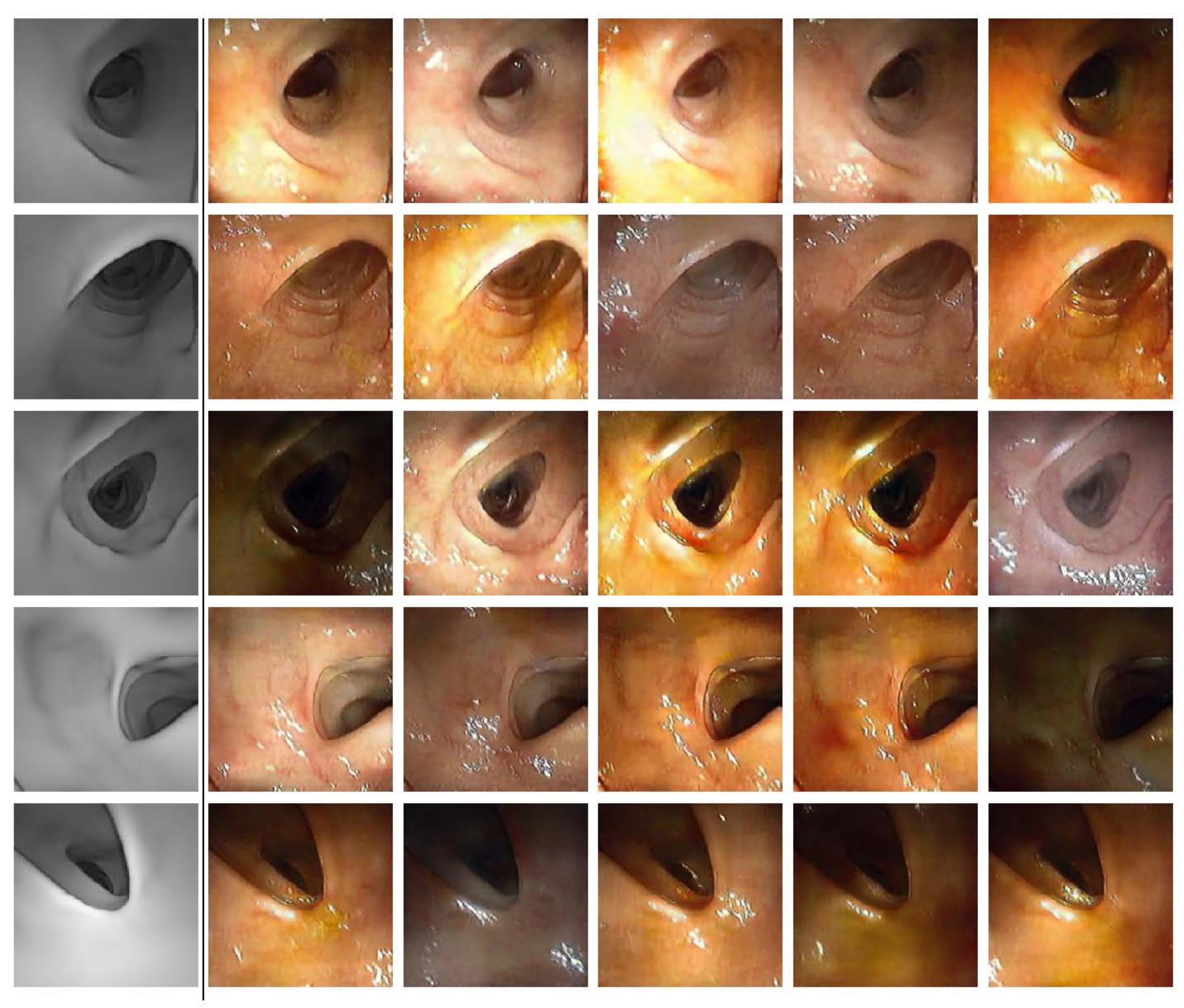}

\caption{This figure depicts our model with a variety of texture and color output for a number of VC input images.}
\vspace{-1em}
\label{fig:supp1}
\end{center}
\end{figure}

\begin{figure}[t!]
\begin{center}

\includegraphics[width=.7\textwidth]{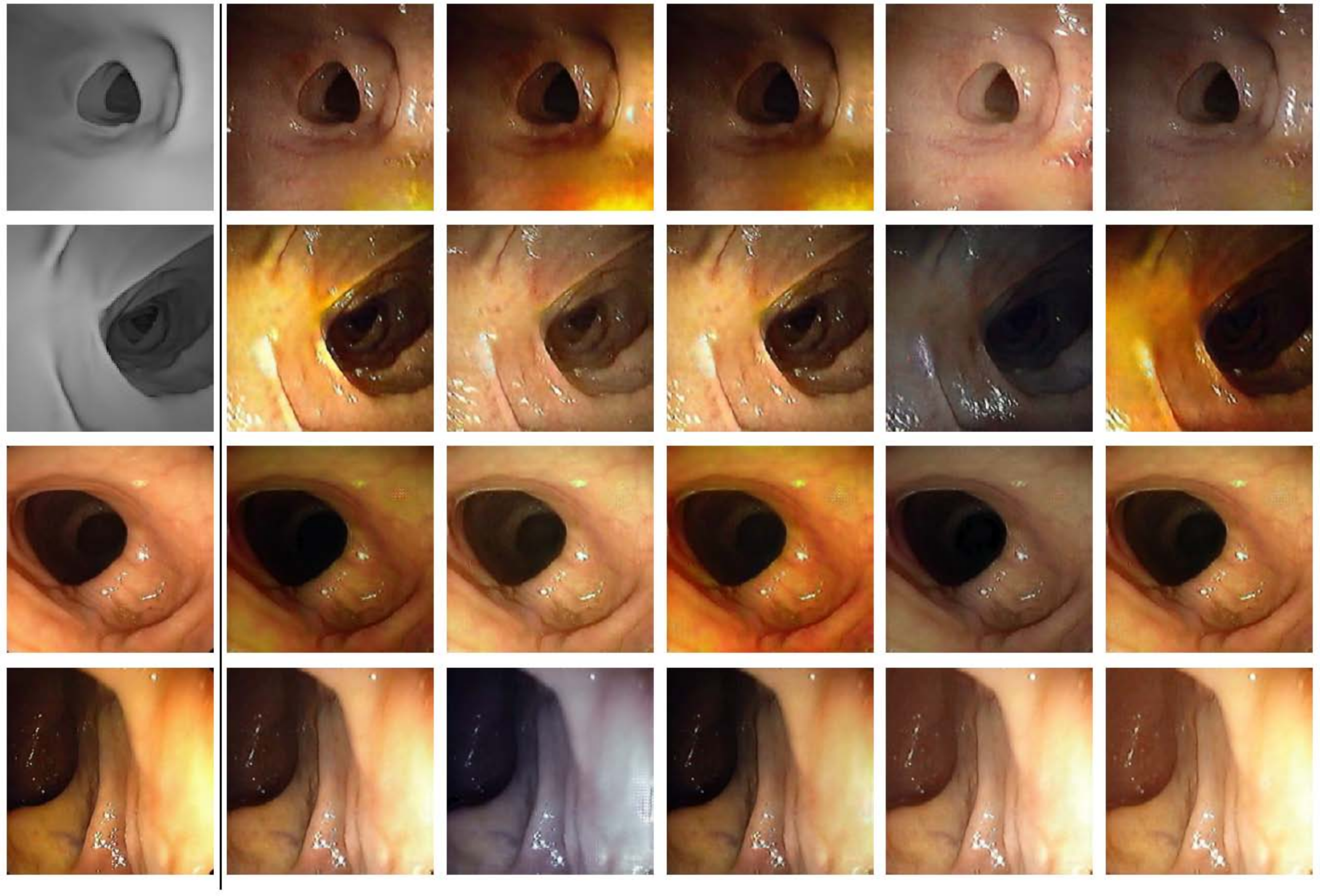}

\caption{This figure depicts our models ability to generate images with different color and lighting. Each image in a row has a different colon-specific color and lighting vector as input. Notice how the specular reflections alter based on the color and lighting vector, but the locations of the reflections remain the same.}
\label{fig:supp2}
\end{center}
\end{figure}

\begin{figure}[t!]
\begin{center}

\includegraphics[width=.7\textwidth]{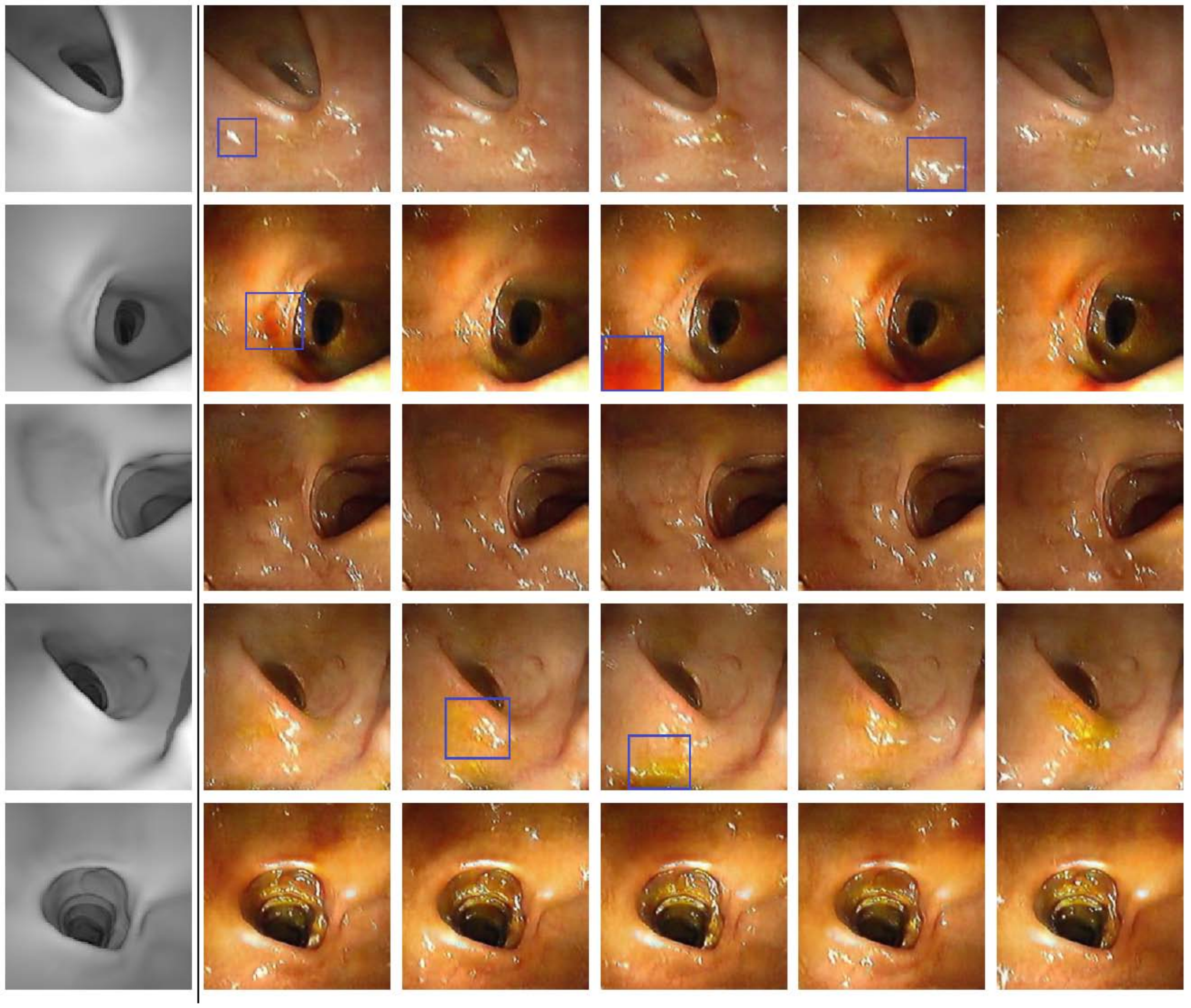}

\caption{This figure depicts our models ability to generate images with different texture and specular reflections using $z_{ts}$. The left most image is CLTS-GAN's input followed its output with a fixed $z_{cl}$ and various $z_{ts}$ vectors. Blue boxes are used to highlight some of the differences.}
\label{fig:supp3}
\end{center}
\end{figure}


\end{document}